\documentclass{article}
\usepackage{arxiv}
\usepackage{amsmath}
\usepackage{amssymb}
\usepackage{graphicx}
\usepackage{hyperref}
\usepackage{enumitem}
\usepackage{listings}
\usepackage{xcolor}

\title{Unexplored Opportunities for Automatic Differentiation in Astrophysics}

\author{
  Marc Bara Iniesta \\
  \texttt{marc.bara.iniesta@gmail.com} \\
  \href{https://orcid.org/0009-0005-1480-5760}{ORCID: 0009-0005-1480-5760} \\
}

\date{July 12th, 2025 \\[0.5em] 
© 2025 Marc Bara – Released under CC BY 4.0 \\[0.5em]
\textit{Patent Pending – U.S. Provisional Patent Application No. 63/842,854}}

\begin{document}

\maketitle

\begin{abstract}
We present a systematic analysis of automatic differentiation (AD) applications in astrophysics, identifying domains where gradient-based optimization could provide significant computational advantages. Building on our previous work with GRAF (Gradient-based Radar Ambiguity Functions), which discovered optimal radar waveforms achieving 4× computational speedup by exploring the trade-off space between conflicting objectives \cite{iniesta2025differentiableradarambiguityfunctions}, we extend this discovery-oriented approach to astrophysical parameter spaces. While AD has been successfully implemented in several areas including gravitational wave parameter estimation and exoplanet atmospheric retrieval, we identify nine astrophysical domains where, to our knowledge, gradient-based exploration methods remain unexplored despite favorable mathematical structure. These opportunities range from discovering novel solutions to the Einstein field equations in exotic spacetime configurations to systematically exploring parameter spaces in stellar astrophysics and planetary dynamics. We present the mathematical foundations for implementing AD in each domain and propose GRASP (Gradient-based Reconstruction of Astrophysical Systems \& Phenomena), a unified framework for differentiable astrophysical computations that transforms traditional optimization problems into systematic exploration of solution spaces. To our knowledge, this is the first work to systematically delineate unexplored domains in astrophysics suitable for automatic differentiation and to provide a unified, mathematically grounded framework (GRASP) to guide their implementation.

\end{abstract}

\section{Introduction}

The application of automatic differentiation (AD) to scientific computing has emerged as a transformative methodology for parameter estimation and optimization problems. Originally developed for neural network training, AD frameworks such as PyTorch \cite{pytorch}, JAX \cite{jax}, and TensorFlow \cite{tensorflow} provide exact gradient computation through computational graphs, enabling efficient optimization in high-dimensional parameter spaces.

In the context of signal processing applications, we previously demonstrated how AD transforms optimization problems into systematic exploration tools. Our GRAF (Gradient-based Radar Ambiguity Functions) system achieved a 4× reduction in computation time by discovering optimal radar waveforms through exploration of the trade-off space between peak sidelobe levels and low probability of intercept characteristics \cite{iniesta2025differentiableradarambiguityfunctions}. Crucially, GRAF demonstrated that gradient-based methods not only optimize faster but discover superior solutions in multi-objective spaces—waveforms that no engineer would have designed yet outperform traditional approaches on all metrics simultaneously. This success in discovering optimal configurations within complex parameter spaces motivated our investigation into similar opportunities in astrophysics.

\subsection{From Optimization to Discovery}

The key insight from GRAF extends beyond computational efficiency: gradient-based exploration of parameter spaces discovers solutions that elude human intuition. In radar waveform design, the joint optimization of conflicting objectives (detection performance vs. stealth characteristics) revealed Pareto-optimal solutions in regions of parameter space that traditional design approaches never explore. This paradigm shift—from optimizing human-conceived designs to discovering the full landscape of optimal solutions—motivates the extension to astrophysical domains.

In astrophysics, many fundamental problems involve exploring vast parameter spaces subject to physical constraints:
\begin{itemize}
\item What spacetime geometries minimize exotic matter while permitting superluminal motion?
\item Which stellar models simultaneously satisfy asteroseismic frequencies and spectroscopic constraints?
\item What antenna gain configurations optimize interferometric calibration across multiple observing conditions?
\end{itemize}

These questions require not just optimization of known solutions, but systematic discovery of all possible solutions within the constraints of physics.

\subsection{The Power of Automatic Differentiation: From Forward Models to Discovery}

The transformative insight underlying both GRAF and GRASP is deceptively simple: if we can compute something, automatic differentiation lets us optimize it through iterative refinement. This principle, familiar from neural network training, extends powerfully to physical systems.

Consider the current workflow in astrophysics when dealing with complex, non-convex optimization landscapes:
\begin{enumerate}
\item \textbf{Forward model}: Given parameters $\theta$, compute observables (e.g., given stellar mass, compute oscillation frequencies)
\item \textbf{Comparison}: Evaluate how well the model matches data or satisfies constraints
\item \textbf{Population-based search}: Use genetic algorithms, particle swarm, or similar methods that require hundreds to thousands of forward evaluations to estimate good parameter updates
\end{enumerate}

These methods work but are computationally expensive because they treat the forward model as a black box, learning about the landscape only through repeated sampling.

Automatic differentiation transforms this into an efficient discovery process:
\begin{enumerate}
\item \textbf{Forward model}: Same as before - compute observables from parameters
\item \textbf{Loss function}: Quantify the mismatch or constraint violation as $\mathcal{L}(\theta)$
\item \textbf{Backward pass}: AD automatically computes $\nabla_\theta \mathcal{L}$ - the exact direction to improve
\item \textbf{Parameter update}: $\theta_{new} = \theta_{old} - \eta \nabla_\theta \mathcal{L}$
\item \textbf{Iterate}: Repeat until convergence
\end{enumerate}

The crucial advantage: while genetic algorithms require hundreds of forward evaluations to estimate a single good update direction, AD provides the exact gradient at a cost comparable to just one or two forward passes. This represents a 100-1000× reduction in computational cost per iteration.

This efficiency enables a paradigm shift. Rather than being limited by computational budget to explore only small regions of parameter space, we can:
\begin{itemize}
\item Start from many random initializations to explore diverse regions efficiently
\item Navigate through million-dimensional parameter spaces (e.g., neural network weights)
\item Discover solutions in regions traditional methods never reach due to computational constraints
\item Find optimal trade-offs between competing objectives in minutes instead of days
\end{itemize}

Just as gradient descent in neural networks discovers feature representations no human would design, gradient-based exploration in physics discovers configurations no scientist would conceive—yet they satisfy all physical laws and optimize our objectives.

\subsection{Technical Foundations of Automatic Differentiation}

While the discovery paradigm drives our approach, we briefly review the mathematical machinery that makes it possible. AD computes exact derivatives by decomposing functions into elementary operations and applying the chain rule systematically.

\textbf{Forward Mode}: For a function $f: \mathbb{R}^n \rightarrow \mathbb{R}^m$, forward mode computes directional derivatives:
\begin{equation}
\dot{y} = \frac{\partial f}{\partial x} \dot{x}
\end{equation}
where $\dot{x}$ is a seed vector. This is efficient when $n \ll m$.

\textbf{Reverse Mode}: More relevant for optimization where typically $m=1$ (scalar loss), reverse mode computes:
\begin{equation}
\bar{x} = \left(\frac{\partial f}{\partial x}\right)^T \bar{y}
\end{equation}
where $\bar{y}$ is the adjoint. This requires one forward pass and one backward pass, regardless of $n$.

\textbf{Complex Differentiation}: For complex-valued functions, the Wirtinger derivatives provide:
\begin{equation}
\frac{\partial f}{\partial z} = \frac{1}{2}\left(\frac{\partial f}{\partial x} - i\frac{\partial f}{\partial y}\right), \quad \frac{\partial f}{\partial z^*} = \frac{1}{2}\left(\frac{\partial f}{\partial x} + i\frac{\partial f}{\partial y}\right)
\end{equation}
where $z = x + iy$. For holomorphic functions, $\frac{\partial f}{\partial z^*} = 0$, recovering standard complex differentiation. This formalism is essential for radio interferometry and other complex-valued problems where functions are generally not holomorphic.

\subsection{Current State of AD in Astrophysics}

To our knowledge, automatic differentiation has already proved useful in a \emph{diverse} set of astrophysical workflows, including gravitational--wave parameter estimation \cite{gw1,gw2}, strong--lensing mass modelling \cite{lens1}, cosmological parameter inference \cite{cosmo1}, exoplanet spectroscopic retrieval \cite{exojax}, pulsar--timing residual analysis \cite{pta1}, transit--photometry fitting \cite{transit1}, fast--radio--burst dispersion modelling \cite{frb1}, Zeeman--Doppler imaging of stellar magnetic fields \cite{zdi1}, and exploratory optimisation of wormhole geometries \cite{worm1}. Collectively, these implementations confirm that gradient--based methods are both \emph{feasible} and \emph{advantageous} whenever the underlying physics is differentiable and the search space is high--dimensional—precisely the regime targeted by \textsc{GRASP}.

Despite these successes, a substantial swath of astrophysical problems with comparable differentiable structure remains unexplored. In this paper we delineate nine such frontiers and provide the mathematical scaffolding required to equip each of them with gradient-based optimisation.

\section{Mathematical Framework for Astrophysical Applications}

\subsection{General Discovery Framework}

Most astrophysical problems can be formulated as exploration of solution spaces subject to constraints:
\begin{equation}
\mathcal{S}^* = \{\theta : \mathcal{C}(\theta) = 0, \mathcal{L}(\theta) \leq \mathcal{L}_{max}\}
\end{equation}
where $\mathcal{S}^*$ represents the set of all valid solutions, $\mathcal{C}(\theta)$ represents physical constraints (e.g., Einstein equations, conservation laws), and $\mathcal{L}(\theta)$ represents objectives to minimize (e.g., exotic matter, fitting residuals).

The discovery process involves:
\begin{enumerate}
\item \textbf{Exploration}: Systematically sampling the parameter space $\theta$
\item \textbf{Constraint satisfaction}: Ensuring physical validity through differentiable constraints
\item \textbf{Pareto frontier identification}: Finding trade-offs between competing objectives
\item \textbf{Solution characterization}: Identifying qualitatively different solution families
\end{enumerate}

This transforms single-point optimization into comprehensive mapping of solution landscapes.

\subsection{Key Differentiable Operations in Astrophysics}

Several mathematical operations appear repeatedly across astrophysical applications:

\textbf{Integration}: Many forward models involve integrals
\begin{equation}
I(\theta) = \int_{\Omega} f(x; \theta) dx
\end{equation}
These can be differentiated using Leibniz's rule when the integrand is smooth.

\textbf{Differential Equations}: Stellar structure, orbital dynamics, and field evolution often involve ODEs/PDEs:
\begin{equation}
\frac{dy}{dt} = f(y, t; \theta), \quad y(0) = y_0(\theta)
\end{equation}
The sensitivity equations or adjoint methods provide gradients.

\textbf{Fourier Transforms}: Essential for interferometry and time series analysis:
\begin{equation}
\tilde{f}(k) = \int f(x) e^{-ikx} dx
\end{equation}
The FFT is differentiable, enabling end-to-end optimization.

\textbf{Matrix Operations}: From Jones calculus to metric tensors:
\begin{equation}
\mathbf{y} = \mathbf{A}(\theta) \mathbf{x}, \quad \text{where } \frac{\partial \mathbf{y}}{\partial \theta} = \frac{\partial \mathbf{A}}{\partial \theta} \mathbf{x}
\end{equation}

\section{Unexplored Applications for Automatic Differentiation}

We now present nine astrophysical domains where, to our knowledge, automatic differentiation has not been applied for systematic exploration of solution spaces. For each application, we demonstrate how gradient-based discovery can reveal solutions beyond human intuition.

\subsection{General Relativistic Applications}

The Einstein field equations and related spacetime metrics constitute inherently differentiable systems, yet optimization approaches for exotic solutions remain largely unexplored.

\subsubsection{Alcubierre Metric Optimization}

The Alcubierre metric \cite{alcubierre1994} represents a solution to the Einstein field equations permitting apparent faster-than-light travel through spacetime distortion:
\begin{equation}
ds^2 = -dt^2 + [dx - v_s f(r_s)dt]^2 + dy^2 + dz^2
\end{equation}
where $f(r_s)$ is the shape function and $r_s = [(x-x_s(t))^2 + y^2 + z^2]^{1/2}$.

The stress-energy tensor components require:
\begin{equation}
T^{00} = \frac{v_s^2}{8\pi} \left[ \frac{(x-x_s)^2}{r_s^2} \left( \frac{df}{dr_s} \right)^2 + f^2 \frac{d^2f}{dr_s^2} \right]
\end{equation}

This expression assumes a static bubble configuration where $\dot{x}_s = 0$. For a moving bubble, additional terms arise from the time dependence of $r_s$ through $x_s(t)$. In our optimization framework, we consider static snapshots where the bubble position is fixed during each gradient computation, making the above expression valid.

\textbf{Discovery Problem}: Explore the landscape of all spacetime geometries that satisfy Einstein's equations while minimizing $\int |T^{\mu\nu}| dV$, subject to boundary conditions $f(0) = 1$, $f(\infty) = 0$.

\paragraph{Differentiable Formulation}

We parameterize the shape function using a neural network or smooth basis functions to ensure differentiability:
\begin{equation}
f(r_s; \theta) = \sum_{k=1}^K \theta_k \phi_k(r_s)
\end{equation}
where $\phi_k$ are basis functions (e.g., radial basis functions, B-splines) and $\theta$ are learnable parameters.

For neural parameterization:
\begin{equation}
f(r_s; \theta) = \sigma(W_L \cdots \sigma(W_2 \sigma(W_1 r_s + b_1) + b_2) \cdots + b_L)
\end{equation}
where $\sigma$ is a smooth activation function and $\theta = \{W_i, b_i\}$. Unlike traditional analytical forms that constrain the search to human-conceived geometries, neural parameterization enables the discovery of novel spacetime configurations that may violate geometric intuition while better satisfying the physical constraints.

\paragraph{Gradient Computation}

The key challenge is computing gradients through the stress-energy tensor. We decompose the computation:

\begin{enumerate}
\item \textbf{Shape function derivatives}: For the basis function parameterization:
   \begin{equation}
   \frac{\partial f}{\partial r_s} = \sum_k \theta_k \frac{\partial \phi_k}{\partial r_s}, \quad \frac{\partial^2 f}{\partial r_s^2} = \sum_k \theta_k \frac{\partial^2 \phi_k}{\partial r_s^2}
   \end{equation}
   For neural network parameterization, we compute these derivatives directly through automatic differentiation of $f(r_s; \theta)$.

\item \textbf{Stress-energy components}: Each component is differentiable:
   \begin{equation}
   \frac{\partial T^{00}}{\partial \theta} = \frac{v_s^2}{8\pi} \left[ \frac{(x-x_s)^2}{r_s^2} \cdot 2\frac{df}{dr_s}\frac{\partial}{\partial \theta}\left(\frac{df}{dr_s}\right) + 2f\frac{\partial f}{\partial \theta}\frac{d^2f}{dr_s^2} + f^2\frac{\partial}{\partial \theta}\left(\frac{d^2f}{dr_s^2}\right) \right]
   \end{equation}

\item \textbf{Volume integral}: We discretize on a spatial grid. For simplicity, we use Cartesian coordinates with uniform spacing:
   \begin{equation}
   E[\theta] = \sum_{i,j,k} |T^{\mu\nu}(x_i, y_j, z_k; \theta)| \Delta x \Delta y \Delta z
   \end{equation}
   Note: For the logarithmic radial spacing mentioned in Eq. 16, appropriate volume weights would be needed.
\end{enumerate}

\paragraph{Implementation Algorithm}

\begin{center}
\fbox{\begin{minipage}{0.9\textwidth}
\textbf{Algorithm: Differentiable Alcubierre Optimization}\\[0.5em]
\textbf{Input:} Initial parameters $\theta_0$, velocity $v_s$, grid resolution $N$\\
\textbf{Output:} Optimized shape function parameters $\theta^*$\\[0.5em]
\begin{enumerate}
\item Initialize spatial grid: $\{x_i, y_j, z_k\}$ for $i,j,k \in [1,N]$
\item \textbf{repeat}
    \begin{enumerate}
    \item[a.] Compute $r_s(i,j,k) = \sqrt{x_i^2 + y_j^2 + z_k^2}$ for all grid points
    \item[b.] Evaluate $f(r_s; \theta)$ using neural network/basis functions
    \item[c.] Compute derivatives $\frac{\partial f}{\partial r_s}$, $\frac{\partial^2 f}{\partial r_s^2}$ via autodiff
    \item[d.] Calculate $T^{\mu\nu}$ components at each grid point
    \item[e.] Compute loss: $\mathcal{L} = \frac{1}{N^3} \sum_{i,j,k} |T^{\mu\nu}(i,j,k)| \Delta x \Delta y \Delta z$
    \item[f.] Add constraints: $\mathcal{L} \gets \mathcal{L} + \lambda_1(f(0) - 1)^2 + \lambda_2 f(r_{max})^2$
    \item[g.] Update: $\theta \gets \theta - \eta \nabla_\theta \mathcal{L}$
    \end{enumerate}
\item \textbf{until} convergence
\end{enumerate}
Note: Weights $\lambda_i$ should be tuned to balance terms (typically $\lambda_1, \lambda_2 \sim 10^2-10^4$ for normalized $T^{\mu\nu}$).
\end{minipage}}
\end{center}

\paragraph{Numerical Considerations}

\begin{itemize}
\item \textbf{Spatial discretization}: The grid must extend sufficiently far to capture the asymptotic behavior. We use logarithmic spacing near the origin:
\begin{equation}
r_i = r_{\min} \exp\left(i \frac{\log(r_{\max}/r_{\min})}{N_r}\right)
\end{equation}

\item \textbf{Boundary conditions}: Enforce constraints using penalty terms:
\begin{equation}
\mathcal{L}_{\text{boundary}} = \lambda_1|f(0) - 1|^2 + \lambda_2|f(r_{\max})|^2 + \lambda_3 \int_0^{r_{\max}} \max(0, -f(r))^2 dr
\end{equation}

\item \textbf{Gradient stability}: The term $(x-x_s)^2/r_s^2$ can cause numerical issues near $r_s = 0$. We regularize:
\begin{equation}
\frac{(x-x_s)^2}{r_s^2} \rightarrow \frac{(x-x_s)^2}{r_s^2 + \epsilon}
\end{equation}
with $\epsilon \sim 10^{-4}$ for single-precision (float32) or $\epsilon \sim 10^{-8}$ for double-precision (float64) computations.

\item \textbf{Efficient implementation}: Vectorize all operations over the spatial grid:
\begin{verbatim}
# Pseudocode
r_s = torch.sqrt(X**2 + Y**2 + Z**2 + eps)
f = shape_network(r_s)
df_dr = torch.autograd.grad(f, r_s, create_graph=True)[0]
d2f_dr2 = torch.autograd.grad(df_dr, r_s, create_graph=True)[0]
T_00 = compute_stress_energy(f, df_dr, d2f_dr2, v_s)
loss = torch.sum(torch.abs(T_00)) * dx * dy * dz
\end{verbatim}
\end{itemize}

\paragraph{Geometry Space Exploration}

The neural parameterization transforms the problem from optimizing a single predetermined geometry to systematically exploring the entire landscape of physically valid spacetime configurations. We can implement several exploration strategies:

\textbf{Multi-start optimization}: Initialize multiple neural networks with different random weights to explore diverse regions of geometry space:
\begin{equation}
\{\theta_i^*\} = \{\arg\min_{\theta_i} E[\theta_i]\}, \quad i = 1, ..., N_{starts}
\end{equation}

\textbf{Geometry generation pipeline}: Train a generative model to produce diverse initial geometries:
\begin{equation}
\theta_{init} \sim p_{\text{generator}}(\theta | \text{constraints})
\end{equation}

\textbf{Population-based exploration}: Combine gradient descent with evolutionary strategies to maintain diversity while optimizing:
\begin{equation}
\theta_i^{(t+1)} = \theta_i^{(t)} - \eta \nabla E[\theta_i^{(t)}] + \epsilon_i^{(t)}
\end{equation}
where $\epsilon_i^{(t)}$ introduces stochastic exploration.

This approach enables discovery of multiple solution families, characterization of the Pareto frontier between competing objectives (e.g., exotic matter vs. bubble size), and identification of qualitatively different geometric configurations that satisfy the same physical constraints. The framework transforms spacetime engineering from a problem of optimizing human-conceived designs to one of discovering the full spectrum of physically realizable geometries.

\subsubsection{Cosmic String Network Evolution}

Cosmic strings \cite{VilenkinShellard} evolve according to the Nambu-Goto action:
\begin{equation}
S = -\mu \int d^2\sigma \sqrt{-\det(\gamma_{ab})}, \quad \gamma_{ab} = \partial_a X^\mu \partial_b X_\mu,
\end{equation}
where $X^\mu(\sigma,\tau)$ describes the string worldsheet in spacetime coordinates $\mu = 0,1,2,3$ and $\sigma^a = (\sigma,\tau)$ are worldsheet coordinates.

The equations of motion derived from (24) are:
\begin{equation}
\partial_a\left(\sqrt{-\gamma} \gamma^{ab} \partial_b X^\mu\right) = 0,
\label{eq:string_eom}
\end{equation}
with $\gamma = \det(\gamma_{ab})$ and $\gamma^{ab}$ the inverse induced metric.

\noindent\textbf{Discovery Problem}: Given observational constraints on the stochastic gravitational wave background $\Omega_{GW}(f)$, find the string network parameters $\theta = (\mu, p_{reconnect}, \xi_{initial})$ that minimize:
\begin{equation}
\mathcal{L}(\theta) = \int \left|\Omega_{GW}^{\text{model}}(f;\theta) - \Omega_{GW}^{\text{obs}}(f)\right|^2 df + \lambda \mathcal{R}(\theta),
\end{equation}
where $\mathcal{R}(\theta)$ penalizes unphysical configurations (e.g., negative string tensions).

\noindent\textbf{Differentiable Formulation}:
\begin{itemize}
\item \textit{Worldsheet Parameterization}: Represent $X^\mu(\sigma,\tau;\theta)$ as a spectral decomposition:
\begin{equation}
X^\mu(\sigma,\tau;\theta) = \sum_{k=1}^K \theta_k^\mu \phi_k(\sigma,\tau),
\end{equation}
where $\phi_k$ are Fourier basis functions for periodic strings or B-splines for non-periodic segments.

\item \textit{Gradient Computation}: The adjoint method computes gradients through the PDE constraint (\ref{eq:string_eom}):
\begin{equation}
\frac{\partial \mathcal{L}}{\partial \theta} = \int \frac{\partial \mathcal{L}}{\partial X^\mu} \frac{\partial X^\mu}{\partial \theta} d^2\sigma + \text{adjoint terms},
\end{equation}
with the adjoint field $\lambda^\mu(\sigma,\tau)$ satisfying:
\begin{equation}
\partial_a\left(\sqrt{-\gamma}\gamma^{ab}\partial_b \lambda^\mu\right) = -\frac{\delta \mathcal{L}}{\delta X^\mu}.
\end{equation}
\end{itemize}

\begin{center}
\fbox{\begin{minipage}{0.9\textwidth}
\textbf{Algorithm: Differentiable Cosmic String Optimization} \\[0.5em]
\textbf{Input:} Initial parameters $\theta_0$, worldsheet resolution $(N_\sigma, N_\tau)$, observational data $\Omega_{GW}^{\text{obs}}(f)$ \\
\textbf{Output:} Optimized parameters $\theta^*$ \\[0.5em]
\begin{enumerate}
\item Initialize adaptive worldsheet grid $(\sigma_i, \tau_j)$
\item \textbf{repeat}
    \begin{enumerate}
    \item[a.] Compute induced metric $\gamma_{ab} = \partial_a X^\mu \partial_b X_\mu$
    \item[b.] Evaluate equation residual $R^\mu = \partial_a(\sqrt{-\gamma}\gamma^{ab}\partial_b X^\mu)$
    \item[c.] Calculate GW spectrum $\Omega_{GW}^{\text{model}}(f;\theta)$
    \item[d.] Compute loss $\mathcal{L} = \|R^\mu\|^2 + \alpha \|\Omega_{GW}^{\text{model}} - \Omega_{GW}^{\text{obs}}\|^2$
    \item[e.] Solve adjoint equation for $\lambda^\mu(\sigma,\tau)$
    \item[f.] Update $\theta \leftarrow \theta - \eta \nabla_\theta\mathcal{L}$
    \end{enumerate}
\item \textbf{until} convergence
\end{enumerate}
\end{minipage}}
\end{center}

\noindent\textbf{Numerical Considerations}:
\begin{itemize}
\item \textbf{Singularity Handling}: The worldsheet metric $\gamma_{ab}$ becomes degenerate at cusps ($\det(\gamma) \to 0$). We regularize with:
\begin{equation}
\sqrt{-\gamma} \to \sqrt{-\gamma + \epsilon}, \quad \epsilon \sim 10^{-8}.
\end{equation}

\item \textbf{Discretization}: Use adaptive mesh refinement near high-curvature regions, with gradient-aware refinement criteria:
\begin{equation}
\Delta \sigma \propto \left\|\frac{\partial \mathcal{L}}{\partial X^\mu}\right\|^{-1/2}.
\end{equation}
\end{itemize}

\noindent\textbf{Exploration Strategies}:
\begin{itemize}
\item \textbf{Multi-Objective Optimization}: Simultaneously minimize $\mathcal{L}(\theta)$ for multiple GW datasets:
\begin{itemize}
\item Pulsar Timing Arrays (PTA) for nanohertz signals
\item LIGO/Virgo/KAGRA for high-frequency bursts
\end{itemize}

\item \textbf{Physics-Informed Initialization}: Seed $\theta$ with values from analytic string network models \cite{VilenkinBook}.
\end{itemize}

\subsubsection{Domain Wall Dynamics}

Domain walls \cite{VilenkinShellard} are topological defects described by the scalar field Lagrangian:
\begin{equation}
\mathcal{L} = \frac{1}{2}\partial_\mu\phi\partial^\mu\phi - V(\phi), \quad V(\phi) = \frac{\lambda}{4}(\phi^2 - \eta^2)^2.
\end{equation}

For a static planar wall at $z=0$, the field equation reduces to:
\begin{equation}
\frac{d^2\phi}{dz^2} = \lambda\phi(\phi^2 - \eta^2), \quad \phi(-\infty) = -\eta, \ \phi(+\infty) = +\eta.
\end{equation}

\textbf{Discovery Problem}: Find field configurations $\phi(z;\theta)$ that minimize the total energy
\begin{equation}
E[\theta] = \int_{-\infty}^\infty \left[ \frac{1}{2}\left(\frac{d\phi}{dz}\right)^2 + V(\phi) \right] dz,
\end{equation}
while satisfying observational constraints from CMB anisotropies \cite{CMBbounds}.

\paragraph{Differentiable Formulation}

Parameterize $\phi(z;\theta)$ using a spectral decomposition:
\begin{equation}
\phi(z;\theta) = \eta \tanh\left(\frac{z}{\delta_0}\right) + \sum_{k=1}^K \theta_k \psi_k(z),
\end{equation}
where $\delta_0 = \sqrt{2/\lambda}\eta^{-1}$ and $\psi_k(z)$ are compactly supported basis functions (e.g., B-splines).

The adjoint field $\xi(z)$ for gradient computation satisfies:
\begin{equation}
\frac{d^2\xi}{dz^2} - \lambda(3\phi^2 - \eta^2)\xi = \frac{\delta E}{\delta \phi}, \quad \xi(\pm\infty) = 0.
\end{equation}

\paragraph{Numerical Considerations}

\begin{itemize}
\item \textit{Grid Construction}: Use coordinate mapping $z \to \tilde{z} = z/(1+|z|)$ to compactify the domain to $[-1,1]$.
\item \textit{Regularization}: Add a small kinetic term to prevent singularities:
\begin{equation}
E \to E + \epsilon \int \left(\frac{d^2\phi}{dz^2}\right)^2 dz, \quad \epsilon \sim 10^{-4}.
\end{equation}
\end{itemize}

\paragraph{Exploration Strategies}

\begin{itemize}
\item \textit{Multi-Objective Optimization}: Simultaneously minimize:
\begin{equation}
\mathcal{L}_1 = E[\theta], \quad \mathcal{L}_2 = \|C_\ell^{\text{model}}(\theta) - C_\ell^{\text{obs}}\|^2.
\end{equation}
\item \textit{Analytic Baseline}: Compare against the exact $\tanh$ solution for verification.
\end{itemize}

Domain walls \cite{VilenkinShellard} are topological defects described by the scalar field Lagrangian:
\begin{equation}
\mathcal{L} = \frac{1}{2}\partial_\mu\phi\partial^\mu\phi - V(\phi), \quad V(\phi) = \frac{\lambda}{4}(\phi^2 - \eta^2)^2.
\end{equation}

For a static planar wall at $z=0$, the field equation reduces to:
\begin{equation}
\frac{d^2\phi}{dz^2} = \lambda\phi(\phi^2 - \eta^2), \quad \phi(-\infty) = -\eta, \ \phi(+\infty) = +\eta.
\end{equation}

\textbf{Discovery Problem}: Find field configurations $\phi(z;\theta)$ that minimize the total energy
\begin{equation}
E[\theta] = \int_{-\infty}^\infty \left[ \frac{1}{2}\left(\frac{d\phi}{dz}\right)^2 + V(\phi) \right] dz,
\end{equation}
while satisfying observational constraints from CMB anisotropies \cite{CMBbounds}.

\paragraph{Differentiable Formulation}

Parameterize $\phi(z;\theta)$ using a spectral decomposition:
\begin{equation}
\phi(z;\theta) = \eta \tanh\left(\frac{z}{\delta_0}\right) + \sum_{k=1}^K \theta_k \psi_k(z),
\end{equation}
where $\delta_0 = \sqrt{2/\lambda}\eta^{-1}$ and $\psi_k(z)$ are compactly supported basis functions (e.g., B-splines).

The adjoint field $\xi(z)$ for gradient computation satisfies:
\begin{equation}
\frac{d^2\xi}{dz^2} - \lambda(3\phi^2 - \eta^2)\xi = \frac{\delta E}{\delta \phi}, \quad \xi(\pm\infty) = 0.
\end{equation}

\paragraph{Numerical Considerations}

\begin{itemize}
\item \textbf{Grid Construction}: Use coordinate mapping $z \to \tilde{z} = z/(1+|z|)$ to compactify the domain to $[-1,1]$.
\item \textbf{Regularization}: Add a small kinetic term to prevent singularities:
\begin{equation}
E \to E + \epsilon \int \left(\frac{d^2\phi}{dz^2}\right)^2 dz, \quad \epsilon \sim 10^{-4}.
\end{equation}
\end{itemize}

\paragraph{Exploration Strategies}

\begin{itemize}
\item \textbf{Multi-Objective Optimization}: Simultaneously minimize:
\begin{equation}
\mathcal{L}_1 = E[\theta], \quad \mathcal{L}_2 = \|C_\ell^{\text{model}}(\theta) - C_\ell^{\text{obs}}\|^2.
\end{equation}
\item \textbf{Analytic Baseline}: Compare against the exact $\tanh$ solution for verification.
\end{itemize}

\subsubsection{False Vacuum Decay and Bubble Nucleation}

The nucleation rate of bubbles in a metastable vacuum \cite{Coleman1977} is determined by the bounce solution minimizing the Euclidean action:
\begin{equation}
S_E = 2\pi^2 \int_0^\infty dr \, r^3 \left[ \frac{1}{2}\left(\frac{d\phi}{dr}\right)^2 + V(\phi) \right]
\end{equation}

The bounce equation derived from $\delta S_E = 0$ is:
\begin{equation}
\frac{d^2\phi}{dr^2} + \frac{3}{r}\frac{d\phi}{dr} = \frac{dV}{d\phi}
\end{equation}
with boundary conditions $\phi(0) = \phi_{\text{top}}$ (field value at the potential barrier) and $\phi(\infty) = \phi_{\text{false}}$ (false vacuum state).

\textbf{Discovery Problem}: Find all field configurations $\phi(r)$ that minimize $S_E$ while satisfying quantum tunneling constraints, determining the decay rate $\Gamma \sim e^{-S_E}$.

\paragraph{Differentiable Formulation}
We parameterize the bounce solution using a neural network with radial symmetry:
\begin{equation}
\phi(r;\theta) = \phi_{\text{false}} + (\phi_{\text{top}} - \phi_{\text{false}})\sigma(W_L[...\sigma(W_1 r + b_1)...] + b_L)
\end{equation}
where $\sigma$ is a smooth activation function (tanh) and $\theta = \{W_i,b_i\}$ are learnable parameters. This automatically satisfies $\phi(0) \approx \phi_{\text{top}}$ through initialization and $\phi(\infty) = \phi_{\text{false}}$ via activation function asymptotics.

\paragraph{Gradient Computation}
The AD workflow requires:
\begin{enumerate}
    \item Forward solution of the ODE (Eq. 2) using the parameterized $\phi(r;\theta)$
    \item Evaluation of $S_E[\theta]$ via adaptive quadrature
    \item Backpropagation through the ODE solver using adjoint methods
    \item Constraint enforcement via penalty terms:
    \begin{equation}
        \mathcal{L} = S_E[\theta] + \lambda_1|\phi(0) - \phi_{\text{top}}|^2 + \lambda_2|\phi(R_{\max}) - \phi_{\text{false}}|^2
    \end{equation}
\end{enumerate}

\begin{center}
\fbox{\begin{minipage}{0.9\textwidth}
\textbf{Algorithm: Differentiable Bubble Optimization}\\[0.5em]
\textbf{Input:} Initial parameters $\theta_0$, potential $V(\phi)$, grid $\{r_i\}_{i=1}^N$\\
\textbf{Output:} Optimized bounce solution $\phi^*(r)$\\[0.5em]
\begin{enumerate}
\item Initialize radial grid with logarithmic spacing $r_i = r_{\min}\exp(i\Delta)$
\item \textbf{repeat}
    \begin{enumerate}
    \item[a.] Compute $\phi(r_i;\theta)$ and derivatives via autodiff
    \item[b.] Evaluate ODE residual $R = \phi'' + \frac{3}{r}\phi' - V'(\phi)$
    \item[c.] Calculate loss $\mathcal{L} = S_E[\theta] + \lambda\|R\|^2$
    \item[d.] Update $\theta \gets \theta - \eta\nabla_\theta\mathcal{L}$
    \end{enumerate}
\item \textbf{until} convergence ($\|\nabla\mathcal{L}\| < \epsilon$)
\end{enumerate}
Note: $\lambda$ should be progressively increased during optimization.
\end{minipage}}
\end{center}

\paragraph{Numerical Considerations}
\begin{itemize}
\item \textbf{Singularity Handling}: Regularize the $1/r$ term:
\begin{equation}
\frac{3}{r} \rightarrow \frac{3}{r + \epsilon}, \quad \epsilon \sim 10^{-8}
\end{equation}

\item \textbf{Adaptive Mesh}: Refine grid where $|dV/d\phi|$ is large:
\begin{equation}
\Delta r_{\text{new}} = \frac{C}{|V'(\phi)| + \delta}
\end{equation}

\item \textbf{Potential Smoothing}: Apply Gaussian filtering if $V(\phi)$ has discontinuities:
\begin{equation}
V_{\text{smooth}} = V \ast \mathcal{N}(0,\sigma^2)
\end{equation}
\end{itemize}

\paragraph{Solution Space Exploration}
\begin{itemize}
\item \textbf{Multi-Field Tunneling}: Extend to $\vec{\phi} = (\phi_1,...,\phi_N)$ with:
\begin{equation}
S_E = 2\pi^2 \int r^3 \left[ \frac{1}{2}\sum_i\left(\frac{d\phi_i}{dr}\right)^2 + V(\vec{\phi}) \right]dr
\end{equation}

\item \textbf{Temperature Effects}: Include finite-$T$ via compact time dimension:
\begin{equation}
S_E^T = S_E + \int_0^\beta d\tau \int d^3x \left[ \frac{1}{2}(\partial_\tau\phi)^2 + V(\phi) \right]
\end{equation}

\item \textbf{Non-Canonical Kinetics}: Generalize for field-space metrics $G_{ij}(\phi)$:
\begin{equation}
S_E = 2\pi^2 \int r^3 \left[ \frac{1}{2}G_{ij}(\phi)\frac{d\phi_i}{dr}\frac{d\phi_j}{dr} + V(\phi) \right]dr
\end{equation}
\end{itemize}

\subsection{Radio Astronomy and Interferometry}

Radio interferometry fundamentally involves complex-valued measurements and calibration challenges that are naturally suited to gradient-based optimization, yet current methods rely on alternating least squares or stochastic sampling without exploiting gradient information.

\subsubsection{Interferometric Calibration}

Radio interferometers measure complex visibilities that encode spatial Fourier components of the sky brightness distribution. The measurement process introduces complex gains that must be calibrated to recover accurate astronomical information \cite{thompson2017}.

The measurement equation for baseline $ij$ is:
\begin{equation}
V_{ij}^{obs}(t,\nu) = g_i(t,\nu) g_j^*(t,\nu) V_{ij}^{true}(t,\nu) + n_{ij}(t,\nu)
\end{equation}
where $g_i$ are complex antenna gains and $n_{ij}$ is thermal noise.

For direction-dependent effects (ionosphere, primary beam), the equation becomes:
\begin{equation}
V_{ij}^{obs} = \sum_{k=1}^{N_{dir}} \mathbf{J}_i^k \mathbf{V}_{ij}^k \mathbf{J}_j^{k\dagger} + \mathbf{n}_{ij}
\end{equation}
where $\mathbf{J}_i^k$ are $2\times2$ Jones matrices and $\mathbf{V}_{ij}^k$ are visibility coherency matrices.

\textbf{Discovery Problem}: Explore the space of calibration solutions that minimize:
\begin{equation}
\mathcal{L}(\theta) = \sum_{ij,t,\nu} \frac{|V_{ij}^{obs} - \mathcal{M}(V_{ij}^{model}; \theta)|^2}{\sigma_{ij}^2} + \mathcal{R}(\theta)
\end{equation}
where $\theta = \{g_i, \phi_{ion}, \mathbf{J}_i^k\}$ includes all calibration parameters and $\mathcal{R}$ enforces physical constraints.

\paragraph{Differentiable Formulation}

We parameterize the complex gains and ionospheric phases using smooth functions:
\begin{equation}
g_i(t,\nu) = a_i(t,\nu) \exp(i\phi_i(t,\nu))
\end{equation}
where amplitudes and phases are parameterized as:
\begin{equation}
a_i(t,\nu) = \exp\left(\sum_{k=0}^{K_a} \theta_{a,k}^i B_k(t) P_k(\nu)\right)
\end{equation}
\begin{equation}
\phi_i(t,\nu) = \sum_{k=0}^{K_\phi} \theta_{\phi,k}^i B_k(t) P_k(\nu) + \phi_{ion}^i(t)
\end{equation}
with $B_k$ being B-spline basis functions in time and $P_k$ polynomials in frequency.

The ionospheric contribution is modeled as a 2D phase screen:
\begin{equation}
\phi_{ion}^i(t) = \frac{8.45 \times 10^9}{\nu} \int_{LOS_i} n_e(x,y,z,t) dl
\end{equation}
We parameterize the electron density using a neural network:
\begin{equation}
n_e(x,y,z,t;\theta_{ion}) = \text{NN}_{\theta_{ion}}(x,y,z,t)
\end{equation}

\paragraph{Gradient Computation}

The complex-valued nature requires Wirtinger derivatives:
\begin{equation}
\frac{\partial \mathcal{L}}{\partial g_i} = \sum_{j} \frac{g_j^* V_{ij}^{model} (V_{ij}^{obs} - g_i g_j^* V_{ij}^{model})^*}{\sigma_{ij}^2}
\end{equation}
\begin{equation}
\frac{\partial \mathcal{L}}{\partial g_i^*} = \sum_{j} \frac{g_j V_{ij}^{model*} (V_{ij}^{obs} - g_i g_j^* V_{ij}^{model})}{\sigma_{ij}^2}
\end{equation}

For the ionospheric screen, we use the chain rule:
\begin{equation}
\frac{\partial \mathcal{L}}{\partial \theta_{ion}} = \sum_i \frac{\partial \mathcal{L}}{\partial \phi_{ion}^i} \frac{\partial \phi_{ion}^i}{\partial n_e} \frac{\partial n_e}{\partial \theta_{ion}}
\end{equation}

\begin{center}
\fbox{\begin{minipage}{0.9\textwidth}
\textbf{Algorithm: Differentiable Interferometric Calibration}\\[0.5em]
\textbf{Input:} Observed visibilities $V_{ij}^{obs}$, sky model $V_{ij}^{model}$, initial $\theta_0$\\
\textbf{Output:} Optimized calibration parameters $\theta^*$\\[0.5em]
\begin{enumerate}
\item Initialize gain amplitudes $a_i = 1$, phases $\phi_i = 0$
\item Precompute basis functions $B_k(t)$, $P_k(\nu)$ on measurement grid
\item \textbf{repeat}
    \begin{enumerate}
    \item[a.] Compute complex gains $g_i = a_i \exp(i\phi_i)$
    \item[b.] Apply gains: $V_{ij}^{cal} = g_i g_j^* V_{ij}^{model}$
    \item[c.] Calculate residuals $R_{ij} = V_{ij}^{obs} - V_{ij}^{cal}$
    \item[d.] Compute loss $\mathcal{L} = \sum_{ij} |R_{ij}|^2/\sigma_{ij}^2$
    \item[e.] Add regularization:
        \begin{itemize}
        \item Smoothness: $\mathcal{L} \gets \mathcal{L} + \lambda_1 \sum_i \|\nabla_t g_i\|^2$
        \item Unity constraint: $\mathcal{L} \gets \mathcal{L} + \lambda_2 \sum_i (|g_i| - 1)^2$
        \end{itemize}
    \item[f.] Compute Wirtinger derivatives $\partial\mathcal{L}/\partial g_i$, $\partial\mathcal{L}/\partial g_i^*$
    \item[g.] Update $\theta \gets \theta - \eta \nabla_\theta \mathcal{L}$
    \end{enumerate}
\item \textbf{until} $\|\nabla \mathcal{L}\| < \epsilon$ or max iterations
\end{enumerate}
\end{minipage}}
\end{center}

\paragraph{Numerical Considerations}

\begin{itemize}
\item \textbf{Gauge Degeneracy}: The product $g_i g_j^*$ is invariant under $g_i \to g_i e^{i\alpha}$. Fix by constraining one reference antenna:
\begin{equation}
g_{ref} = |g_{ref}|, \quad \text{Im}(g_{ref}) = 0
\end{equation}

\item \textbf{Frequency Dependence}: For wide bandwidths, model frequency-dependent gains:
\begin{equation}
g_i(\nu) = g_i^0 \left(\frac{\nu}{\nu_0}\right)^{\alpha_i} \exp\left(i\tau_i(\nu - \nu_0)\right)
\end{equation}

\item \textbf{Outlier Robustness}: Replace $L_2$ loss with Huber loss:
\begin{equation}
\rho(x) = \begin{cases}
\frac{1}{2}x^2 & |x| \leq \delta \\
\delta(|x| - \frac{1}{2}\delta) & |x| > \delta
\end{cases}
\end{equation}
\end{itemize}

\paragraph{Solution Space Exploration}

\begin{itemize}
\item \textbf{Multi-Scale Optimization}: Hierarchical approach from coarse to fine time/frequency resolution:
\begin{equation}
\theta^{(k+1)} = \arg\min_\theta \mathcal{L}(\theta; \Delta t_k, \Delta\nu_k), \quad \Delta t_{k+1} = \Delta t_k/2
\end{equation}

\item \textbf{Direction-Dependent Calibration}: Parameterize Jones matrices using quaternions for $SU(2)$ constraint:
\begin{equation}
\mathbf{J} = q_0 \mathbf{I} + i(q_1 \sigma_x + q_2 \sigma_y + q_3 \sigma_z), \quad \sum_i q_i^2 = 1
\end{equation}

\item \textbf{Joint Sky-Calibration}: Simultaneously optimize sky model and calibration:
\begin{equation}
\min_{I_{sky}, \theta_{cal}} \sum_{ij} \frac{|V_{ij}^{obs} - g_i g_j^* \mathcal{F}(I_{sky})|^2}{\sigma_{ij}^2} + \lambda TV(I_{sky})
\end{equation}
\end{itemize}

\subsubsection{Black Hole Image Reconstruction}

Very Long Baseline Interferometry (VLBI) observations of black holes \cite{eht2019} present an extreme sparse sampling problem in the Fourier domain. The Event Horizon Telescope's success relied on regularized maximum likelihood methods that could benefit from gradient-based exploration of the image and hyperparameter space.

The forward model relates the image $I(x,y)$ to observed visibilities:
\begin{equation}
V(u,v) = \int\int I(x,y) \exp[-2\pi i(ux + vy)] dx dy
\end{equation}
where $(u,v)$ are spatial frequencies sampled by the interferometer.

For black hole imaging, relativistic effects modify the relationship:
\begin{equation}
I_{obs}(\alpha,\beta) = \int_{\gamma} j_\nu g^3 e^{-\tau_\nu} d\lambda
\end{equation}
where the integral follows null geodesics $\gamma$ in the Kerr metric, $j_\nu$ is the emission coefficient, $g$ is the gravitational redshift factor, and $\tau_\nu$ is the optical depth.

\textbf{Discovery Problem}: Explore the space of physically valid black hole images that satisfy:
\begin{equation}
\mathcal{L}(I, \lambda) = \chi^2_{data}(I) + \sum_{k} \lambda_k \mathcal{R}_k(I)
\end{equation}
where $\mathcal{R}_k$ are regularizers encoding physical priors.

\paragraph{Differentiable Formulation}

We parameterize the image using a continuous representation:
\begin{equation}
I(x,y;\theta) = \sum_{k=1}^K \theta_k \phi_k(x,y) + I_{NN}(x,y;\theta_{NN})
\end{equation}
where $\phi_k$ are fixed basis functions (e.g., shapelets, wavelets) and $I_{NN}$ is a neural network component for capturing fine structure.

The data fidelity term with closure quantities:
\begin{equation}
\chi^2_{data} = \chi^2_{vis} + \chi^2_{amp} + \chi^2_{cphase} + \chi^2_{logcamp}
\end{equation}

For closure phases (which cancel station-based errors):
\begin{equation}
\chi^2_{cphase} = \sum_{ijk} \frac{|\arg(V_{ij}V_{jk}V_{ki}) - \arg(\tilde{V}_{ij}\tilde{V}_{jk}\tilde{V}_{ki})|^2}{\sigma_{cphase}^2}
\end{equation}

\paragraph{Gradient Computation}

The key challenge is differentiating through the Fourier transform and closure quantities:

\begin{enumerate}
\item \textbf{Fourier Transform}: Using the FFT with appropriate padding:
\begin{equation}
\frac{\partial \mathcal{L}}{\partial I} = \mathcal{F}^{-1}\left[\frac{\partial \mathcal{L}}{\partial \tilde{V}}\right]
\end{equation}

\item \textbf{Closure Phases}: Require careful handling of phase wrapping:
\begin{equation}
\frac{\partial \chi^2_{cphase}}{\partial V_{ij}} = \sum_{k} \frac{2(\Delta\phi_{ijk})}{\sigma_{ijk}^2} \frac{\partial \arg(V_{ij}V_{jk}V_{ki})}{\partial V_{ij}}
\end{equation}
where the argument derivative uses:
\begin{equation}
\frac{\partial \arg(z)}{\partial z} = \frac{1}{2i}\left(\frac{1}{z} - \frac{1}{z^*}\right)
\end{equation}

\item \textbf{Ray Tracing}: For Kerr metric ray tracing, use automatic differentiation through the geodesic equations:
\begin{equation}
\frac{d^2 x^\mu}{d\lambda^2} + \Gamma^\mu_{\alpha\beta} \frac{dx^\alpha}{d\lambda} \frac{dx^\beta}{d\lambda} = 0
\end{equation}
\end{enumerate}

\begin{center}
\fbox{\begin{minipage}{0.9\textwidth}
\textbf{Algorithm: Differentiable Black Hole Imaging}\\[0.5em]
\textbf{Input:} Visibility data $\{V_{ij}^{obs}\}$, initial image $I_0$, regularization weights $\{\lambda_k\}$\\
\textbf{Output:} Reconstructed image $I^*$ and optimized hyperparameters $\lambda^*$\\[0.5em]
\begin{enumerate}
\item Initialize image with Gaussian blob: $I_0(x,y) = A\exp(-(x^2+y^2)/2\sigma^2)$
\item Precompute $(u,v)$ coverage and noise statistics
\item \textbf{repeat}
    \begin{enumerate}
    \item[a.] Compute model visibilities: $\tilde{V} = \mathcal{F}(I)$
    \item[b.] Interpolate to observed $(u,v)$ points using differentiable kernel
    \item[c.] Calculate closure quantities from model
    \item[d.] Compute data terms:
        \begin{itemize}
        \item $\chi^2_{vis} = \sum_{ij} |V_{ij}^{obs} - \tilde{V}_{ij}|^2/\sigma_{ij}^2$
        \item $\chi^2_{cphase} = \sum_{ijk} (\phi_{ijk}^{obs} - \tilde{\phi}_{ijk})^2/\sigma_{\phi}^2$
        \end{itemize}
    \item[e.] Add regularization terms:
        \begin{itemize}
        \item Total variation: $TV(I) = \sum_{x,y} \sqrt{(\nabla_x I)^2 + (\nabla_y I)^2 + \epsilon}$
        \item Entropy: $S(I) = -\sum_{x,y} I \log(I/I_{prior})$
        \item Ring prior: $R(I) = \sum_{x,y} I(x,y) K_{ring}(x,y)$
        \end{itemize}
    \item[f.] Update image: $I \gets I - \eta_I \nabla_I \mathcal{L}$
    \item[g.] Update hyperparameters: $\lambda \gets \lambda - \eta_\lambda \nabla_\lambda \mathcal{L}$
    \end{enumerate}
\item \textbf{until} convergence
\end{enumerate}
\end{minipage}}
\end{center}

\paragraph{Numerical Considerations}

\begin{itemize}
\item \textbf{Positivity Constraint}: Enforce $I \geq 0$ using projected gradient descent:
\begin{equation}
I_{k+1} = \max(0, I_k - \eta \nabla \mathcal{L})
\end{equation}
or parameterize $I = \exp(\tilde{I})$ to ensure positivity.

\item \textbf{Multi-Resolution}: Use wavelet decomposition for efficient large-scale optimization:
\begin{equation}
I(x,y) = \sum_{j,k,l} c_{jkl} \psi_{jkl}(x,y)
\end{equation}
where $\psi_{jkl}$ are wavelet basis functions at scale $j$.

\item \textbf{Fourier Interpolation}: Use Kaiser-Bessel kernels for accurate gradients:
\begin{equation}
\tilde{V}(u,v) = \sum_{m,n} I_{m,n} K(u - u_m, v - v_n)
\end{equation}
\end{itemize}

\paragraph{Exploration Strategies}

\begin{itemize}
\item \textbf{Multi-Frequency Synthesis}: Joint reconstruction across observing bands:
\begin{equation}
\mathcal{L} = \sum_{\nu} w_\nu \chi^2_{data}(I_\nu) + \lambda_{spec} \sum_{x,y} \|\nabla_\nu I(x,y,\nu)\|^2
\end{equation}

\item \textbf{Polarimetric Imaging}: Extend to full Stokes parameters:
\begin{equation}
\mathbf{V}_{ij} = \begin{pmatrix} V_{I} + V_{Q} & V_{U} + iV_{V} \\ V_{U} - iV_{V} & V_{I} - V_{Q} \end{pmatrix}_{ij}
\end{equation}

\item \textbf{Dynamic Imaging}: For time-variable sources:
\begin{equation}
\min_{I(x,y,t)} \sum_t \chi^2_{data}(I_t) + \lambda_{temp} \int |\partial_t I|^2 dt
\end{equation}

\item \textbf{Bayesian Hyperparameter Selection}: Optimize marginal likelihood:
\begin{equation}
\lambda^* = \arg\max_\lambda \log p(V^{obs}|\lambda) = \arg\max_\lambda \left[-\frac{1}{2}\mathbf{v}^T C^{-1} \mathbf{v} - \frac{1}{2}\log|C|\right]
\end{equation}
where $C = C_{data} + C_{prior}(\lambda)$.
\end{itemize}

The differentiable formulation enables systematic exploration of the image space, discovering features that traditional methods might miss due to local minima, while simultaneously optimizing regularization parameters that balance data fidelity against physical priors. This transforms black hole imaging from an art requiring manual hyperparameter tuning to a principled discovery process.

\subsection{Stellar and Planetary Astrophysics}

Stellar and planetary systems present optimization problems spanning vast scales—from stellar cores where nuclear fusion occurs to planetary surfaces where thermal forces alter orbits. Despite smooth parameter dependencies and well-posed forward models, these domains remain largely unexplored by gradient-based methods.

\subsubsection{Asteroseismic Inversion}

Stars pulsate in global oscillation modes that probe their internal structure \cite{aerts2010}. These acoustic and gravity waves satisfy the stellar pulsation equations, with frequencies encoding information about density stratification, sound speed profiles, and rotation. Current inversion methods rely on grid searches or genetic algorithms without exploiting gradient information.

The stellar oscillation equations in the adiabatic approximation are:
\begin{equation}
\frac{d\xi_r}{dr} = \left(\frac{2}{r} - \frac{1}{\Gamma_1}\frac{d\ln p}{dr}\right)\xi_r + \frac{1}{\Gamma_1 p}\left(\frac{S_\ell^2}{\omega^2} - 1\right)P'
\end{equation}
\begin{equation}
\frac{dP'}{dr} = \rho\left(\omega^2 - N^2\right)\xi_r + \frac{d\ln p}{dr}P'
\end{equation}
where $\xi_r$ is the radial displacement, $P'$ is the Eulerian pressure perturbation, $S_\ell$ is the Lamb frequency, and $N$ is the Brunt-Väisälä frequency.

The eigenvalue problem $\mathcal{L}\xi = \omega^2\xi$ yields discrete frequencies $\{\omega_{n\ell}\}$ that depend on stellar structure parameters $\theta = \{M, R, Y_0, Z, \alpha_{MLT}, f_{ov}, ...\}$.

\textbf{Discovery Problem}: Given observed frequencies $\nu_{n\ell}^{obs}$ with uncertainties $\sigma_{n\ell}$, explore the space of stellar models that minimize:
\begin{equation}
\mathcal{L}(\theta) = \sum_{n,\ell} \frac{(\nu_{n\ell}^{obs} - \nu_{n\ell}^{model}(\theta))^2}{\sigma_{n\ell}^2} + \sum_k \lambda_k \mathcal{R}_k(\theta)
\end{equation}
where $\mathcal{R}_k$ encode constraints from spectroscopy, evolutionary tracks, and physical consistency.

\paragraph{Differentiable Formulation}

The key challenge is differentiating through the eigenvalue problem. We use the Hellmann-Feynman theorem:
\begin{equation}
\frac{\partial \omega_{n\ell}^2}{\partial \theta_i} = \langle \xi_{n\ell} | \frac{\partial \mathcal{L}}{\partial \theta_i} | \xi_{n\ell} \rangle
\end{equation}
where $|\xi_{n\ell}\rangle$ is the normalized eigenfunction.

For the stellar structure, we solve the equations of hydrostatic equilibrium:
\begin{equation}
\frac{dm}{dr} = 4\pi r^2 \rho, \quad \frac{dP}{dr} = -\frac{Gm\rho}{r^2}, \quad \frac{dL}{dr} = 4\pi r^2 \rho \epsilon
\end{equation}
\begin{equation}
\frac{dT}{dr} = -\frac{3\kappa \rho L}{16\pi acT^3 r^2} \quad \text{(radiative)}, \quad \frac{dT}{dr} = \left(1 - \frac{1}{\gamma}\right)\frac{T}{P}\frac{dP}{dr} \quad \text{(convective)}
\end{equation}

These ODEs are differentiable, allowing gradients via the adjoint method:
\begin{equation}
\frac{\partial \mathcal{L}}{\partial \theta} = \int_0^R \lambda^T(r) \frac{\partial f}{\partial \theta}(y(r), r, \theta) dr
\end{equation}
where $\lambda(r)$ is the adjoint variable satisfying a backward ODE.

\paragraph{Gradient Computation}

We decompose the gradient calculation into three stages:

\begin{enumerate}
\item \textbf{Structure Derivatives}: For parameter $\theta_i$, solve the linearized structure equations:
\begin{equation}
\frac{d}{dr}\left(\frac{\partial y}{\partial \theta_i}\right) = \frac{\partial f}{\partial y}\frac{\partial y}{\partial \theta_i} + \frac{\partial f}{\partial \theta_i}
\end{equation}
where $y = (m, P, L, T)^T$ and $f$ represents the RHS of the structure equations.

\item \textbf{Eigenvalue Derivatives}: With perturbed structure, compute:
\begin{equation}
\frac{\partial \omega_{n\ell}^2}{\partial \theta_i} = \int_0^R \xi_r^2 \left[\frac{\partial}{\partial \theta_i}(\rho \omega^2 - A)\right] dr
\end{equation}
where $A$ contains the coefficient functions in the pulsation equations.

\item \textbf{Surface Correction}: Account for near-surface errors:
\begin{equation}
\nu_{n\ell}^{corrected} = \nu_{n\ell}^{model} + a_0 \left(\frac{\nu_{n\ell}}{\nu_{max}}\right)^{a_1}
\end{equation}
with $a_0, a_1$ as additional parameters.
\end{enumerate}

\begin{center}
\fbox{\begin{minipage}{0.9\textwidth}
\textbf{Algorithm: Differentiable Asteroseismic Inversion}\\[0.5em]
\textbf{Input:} Observed frequencies $\{\nu_{n\ell}^{obs}\}$, spectroscopic constraints $(T_{eff}, \log g, [Fe/H])$\\
\textbf{Output:} Stellar parameters $\theta^*$ and internal structure\\[0.5em]
\begin{enumerate}
\item Initialize with scaling relations: $M \approx (\nu_{max}/\nu_{max,\odot})^3(\Delta\nu/\Delta\nu_\odot)^{-4}M_\odot$
\item Construct initial model using 1D stellar evolution code
\item \textbf{repeat}
    \begin{enumerate}
    \item[a.] Solve structure equations $(m,P,L,T)(r)$ via shooting method
    \item[b.] Compute coefficient functions $A(r), V(r), U(r)$ for pulsation
    \item[c.] Solve eigenvalue problem for frequencies $\{\omega_{n\ell}\}$
    \item[d.] Apply surface correction and mode identification
    \item[e.] Compute seismic $\chi^2$:
        $$\chi^2_{seis} = \sum_{n,\ell} \frac{(\nu_{n\ell}^{obs} - \nu_{n\ell}^{model})^2}{\sigma_{n\ell}^2}$$
    \item[f.] Add spectroscopic constraints:
        $$\chi^2_{spec} = \frac{(T_{eff}^{obs} - T_{eff}^{model})^2}{\sigma_{T}^2} + \frac{(\log g^{obs} - \log g^{model})^2}{\sigma_{\log g}^2}$$
    \item[g.] Solve adjoint equations backward from surface
    \item[h.] Compute gradients via adjoint method
    \item[i.] Update: $\theta \gets \theta - \eta \nabla_\theta(\chi^2_{seis} + \lambda\chi^2_{spec})$
    \end{enumerate}
\item \textbf{until} convergence or frequency residuals $< 0.1\mu$Hz
\end{enumerate}
\end{minipage}}
\end{center}

\paragraph{Numerical Considerations}

\begin{itemize}
\item \textbf{Mode Identification}: Use échelle diagram topology:
\begin{equation}
\nu_{n\ell} \approx \Delta\nu(n + \frac{\ell}{2} + \epsilon) + \delta\nu_{n\ell}
\end{equation}
Enforce ordering constraints: $\nu_{n+1,\ell} > \nu_{n,\ell}$.

\item \textbf{Adaptive Mesh}: Concentrate grid points near discontinuities:
\begin{equation}
\Delta r \propto \left|\frac{d^2\rho}{dr^2}\right|^{-1/3}
\end{equation}

\item \textbf{Regularization}: Penalize unphysical gradients:
\begin{equation}
\mathcal{R}_{smooth} = \int_0^R \left|\frac{d^2 \ln \rho}{dr^2}\right|^2 dr
\end{equation}
\end{itemize}

\paragraph{Exploration Strategies}

\begin{itemize}
\item \textbf{Glitch Signatures}: Parameterize sharp features:
\begin{equation}
\delta c^2(r) = A_{BCZ} \exp\left(-\frac{(r-r_{BCZ})^2}{2w^2}\right) + A_{He} \delta(r - r_{He})
\end{equation}
where $r_{BCZ}$ and $r_{He}$ locate the base of convection zone and helium ionization.

\item \textbf{Rotation Inversion}: Include rotational splitting:
\begin{equation}
\nu_{n\ell m} = \nu_{n\ell 0} + m \int_0^R K_{n\ell}(r) \Omega(r) dr
\end{equation}
Parameterize $\Omega(r)$ with splines and jointly optimize.

\item \textbf{Ensemble Analysis}: For cluster members with common age/composition:
\begin{equation}
\mathcal{L}_{ensemble} = \sum_{i=1}^{N_{stars}} \chi^2_i + \lambda_{age} \sum_{i<j} (t_i - t_j)^2 + \lambda_{Z} \sum_{i<j} (Z_i - Z_j)^2
\end{equation}
\end{itemize}

\subsubsection{Radial Velocity Exoplanet Characterization}

The radial velocity method \cite{mayor1995} detects exoplanets through stellar wobble induced by gravitational interaction. Despite being mathematically smooth with well-defined derivatives, current analyses use Markov Chain Monte Carlo or genetic algorithms without gradient exploitation.

The stellar radial velocity due to $N_p$ planets is:
\begin{equation}
v_r(t) = \sum_{p=1}^{N_p} K_p[\cos(f_p(t) + \omega_p) + e_p\cos(\omega_p)] + v_{sys} + v_{inst}(t)
\end{equation}
where $K_p$ is the semi-amplitude, $f_p$ is the true anomaly, $\omega_p$ is the argument of periastron, and $e_p$ is eccentricity.

The true anomaly relates to time through Kepler's equation:
\begin{equation}
M = E - e\sin E, \quad M = \frac{2\pi}{P}(t - T_0)
\end{equation}
where $M$ is mean anomaly and $E$ is eccentric anomaly.

\textbf{Discovery Problem}: Explore the parameter space $\theta = \{K_p, P_p, e_p, \omega_p, T_{0,p}\}_{p=1}^{N_p}$ that minimizes:
\begin{equation}
\mathcal{L}(\theta) = \sum_{i=1}^{N_{obs}} \frac{(v_r^{obs}(t_i) - v_r^{model}(t_i;\theta))^2}{\sigma_i^2} + \mathcal{R}(\theta)
\end{equation}
discovering both the number of planets and their orbital configurations.

\paragraph{Differentiable Formulation}

The key challenge is differentiating through Kepler's equation. We solve it iteratively:
\begin{equation}
E_{n+1} = M + e\sin E_n
\end{equation}
converging to tolerance $|E_{n+1} - E_n| < \epsilon$.

For gradients, we use implicit differentiation. At convergence, $E$ satisfies:
\begin{equation}
g(E, M, e) = E - e\sin E - M = 0
\end{equation}

By the implicit function theorem:
\begin{equation}
\frac{\partial E}{\partial M} = \frac{1}{1 - e\cos E}, \quad \frac{\partial E}{\partial e} = \frac{\sin E}{1 - e\cos E}
\end{equation}

The true anomaly is:
\begin{equation}
f = 2\arctan\left(\sqrt{\frac{1+e}{1-e}}\tan\frac{E}{2}\right)
\end{equation}
with derivatives following from the chain rule.

\paragraph{Gradient Computation}

For each orbital parameter, we trace gradients through the complete model:

\begin{enumerate}
\item \textbf{Period Gradient}:
\begin{equation}
\frac{\partial v_r}{\partial P} = K\frac{\partial}{\partial P}[\cos(f + \omega)] = -K\sin(f + \omega)\frac{\partial f}{\partial P}
\end{equation}
where:
\begin{equation}
\frac{\partial f}{\partial P} = \frac{\partial f}{\partial E}\frac{\partial E}{\partial M}\frac{\partial M}{\partial P} = \frac{\partial f}{\partial E}\frac{1}{1-e\cos E}\frac{-2\pi(t-T_0)}{P^2}
\end{equation}

\item \textbf{Eccentricity Gradient}: More complex due to appearance in multiple terms:
\begin{equation}
\frac{\partial v_r}{\partial e} = K\cos\omega + K\frac{\partial}{\partial e}[\cos(f + \omega)]
\end{equation}

\item \textbf{Multi-Planet Coupling}: Account for gravitational interactions:
\begin{equation}
\mathcal{L}_{stability} = \sum_{i<j} \exp\left(-\frac{\Delta a_{ij}}{R_{Hill}}\right)
\end{equation}
\end{enumerate}

\begin{center}
\fbox{\begin{minipage}{0.9\textwidth}
\textbf{Algorithm: Differentiable RV Exoplanet Discovery}\\[0.5em]
\textbf{Input:} RV measurements $\{t_i, v_i, \sigma_i\}$, stellar mass $M_*$\\
\textbf{Output:} Number of planets $N_p^*$ and orbital parameters $\theta^*$\\[0.5em]
\begin{enumerate}
\item Initialize with periodogram peaks: identify $N_{init}$ candidate periods
\item For $N_p = 1$ to $N_{max}$:
    \begin{enumerate}
    \item[a.] Initialize orbital parameters for $N_p$ planets
    \item[b.] \textbf{repeat}
        \begin{enumerate}
        \item Compute mean anomalies: $M_p(t_i) = 2\pi(t_i - T_{0,p})/P_p$
        \item Solve Kepler's equation for $E_p(t_i)$ via Newton's method
        \item Calculate true anomalies: $f_p(t_i)$ from $E_p(t_i)$
        \item Evaluate model: $v_r^{model}(t_i) = \sum_p K_p[\cos(f_p + \omega_p) + e_p\cos\omega_p]$
        \item Compute residuals and $\chi^2$
        \item Add regularization:
            - Eccentricity prior: $\mathcal{R}_e = \sum_p \beta e_p^2$
            - Stability constraint: $\mathcal{R}_{stab}$
        \item Compute gradients via automatic differentiation
        \item Update: $\theta \gets \theta - \eta\nabla_\theta\mathcal{L}$
        \end{enumerate}
    \item[c.] \textbf{until} convergence
    \item[d.] Compute BIC: $BIC = \chi^2 + k\ln N_{obs}$ where $k = 5N_p$
    \end{enumerate}
\item Select $N_p^*$ minimizing BIC
\item Refine with activity modeling: $v_{activity}(t) = \sum_k A_k\sin(2\pi t/P_{rot} + \phi_k)$
\end{enumerate}
\end{minipage}}
\end{center}

\paragraph{Numerical Considerations}

\begin{itemize}
\item \textbf{Kepler Solver Stability}: Use Markley's method for $e > 0.95$:
\begin{equation}
E = M + e\sin M \frac{1 + e\cos M}{1 - e^2 + e\cos M}
\end{equation}

\item \textbf{Period Aliasing}: Search periods on logarithmic grid:
\begin{equation}
P_k = P_{min} \exp\left(k \frac{\ln(P_{max}/P_{min})}{N_{grid}}\right)
\end{equation}

\item \textbf{Stellar Activity}: Model quasi-periodic variations:
\begin{equation}
v_{GP}(t) \sim \mathcal{GP}(0, k(t,t')) \text{ where } k = A\exp\left(-\frac{|t-t'|^2}{2l^2} - \Gamma\sin^2\frac{\pi|t-t'|}{P_{rot}}\right)
\end{equation}
\end{itemize}

\paragraph{Exploration Strategies}

\begin{itemize}
\item \textbf{Hierarchical Discovery}: Start with strongest signal, progressively add planets:
\begin{equation}
\theta_{N_p+1} = \{\theta_{N_p}^*, \theta_{new}\}
\end{equation}

\item \textbf{Transit-RV Joint Fit}: For transiting systems:
\begin{equation}
\mathcal{L}_{joint} = \chi^2_{RV} + \chi^2_{transit} + \lambda|\rho_* - \rho_{model}|^2
\end{equation}
where stellar density $\rho_*$ links both datasets.

\item \textbf{Population-Level Inference}: Hierarchical Bayesian approach:
\begin{equation}
\theta_i \sim p(\theta|\phi), \quad \phi \sim p(\phi)
\end{equation}
where $\phi$ are hyperparameters describing the population.
\end{itemize}

\subsubsection{Asteroid Thermophysical Characterization}

The Yarkovsky effect \cite{bottke2006} causes orbital drift in asteroids due to anisotropic thermal emission. This subtle force depends on thermal inertia, rotation state, and surface properties—parameters currently constrained through computationally expensive Monte Carlo methods.

The surface temperature \(T(\theta,\phi,t)\) satisfies the heat–diffusion equation:
\[
  \frac{\partial T}{\partial t} \;=\; \kappa \nabla^{2} T,
\]

\begin{equation}
\rho C_p \frac{\partial T}{\partial t} = \frac{1}{r^2}\frac{\partial}{\partial r}\left(r^2 \kappa \frac{\partial T}{\partial r}\right)
\end{equation}
with boundary condition:
\begin{equation}
\kappa\frac{\partial T}{\partial r}\bigg|_{r=R} = (1-A)F_\odot \cos\xi - \epsilon\sigma T^4
\end{equation}
where $\xi$ is the solar zenith angle and $F_\odot$ is the solar flux.

The recoil acceleration from thermal emission is:
\begin{equation}
\mathbf{a}_{Yark} = -\frac{2}{3mc} \int_S \epsilon\sigma T^4 \hat{n} dS
\end{equation}

Over many orbits, this produces secular drift:
\begin{equation}
\frac{da}{dt} = \frac{2}{na}\mathbf{a}_{Yark} \cdot \hat{t}
\end{equation}
where $\hat{t}$ is the transverse direction.

\textbf{Discovery Problem}: Given observed orbital drift $(da/dt)_{obs}$ from astrometry, discover thermal and physical parameters $\theta = \{\Gamma, \rho, C_p, A, \epsilon, P_{rot}, \gamma\}$ that minimize:
\begin{equation}
\mathcal{L}(\theta) = \left(\frac{(da/dt)_{obs} - (da/dt)_{model}(\theta)}{\sigma_{da/dt}}\right)^2 + \sum_k \lambda_k \mathcal{R}_k(\theta)
\end{equation}

\paragraph{Differentiable Formulation}

We discretize the heat equation using finite differences in depth and spherical harmonics on the surface:
\begin{equation}
T(r, \theta, \phi, t) = \sum_{\ell=0}^{L_{max}} \sum_{m=-\ell}^{\ell} T_{\ell m}(r, t) Y_{\ell m}(\theta, \phi)
\end{equation}

The time evolution becomes:
\begin{equation}
\frac{\partial T_{\ell m}}{\partial t} = \frac{\kappa}{\rho C_p r^2}\frac{\partial}{\partial r}\left(r^2 \frac{\partial T_{\ell m}}{\partial r}\right)
\end{equation}

For the surface heating, we expand:
\begin{equation}
\cos\xi = \sum_{\ell,m} S_{\ell m}(t) Y_{\ell m}(\theta, \phi)
\end{equation}
where $S_{\ell m}$ encode the solar direction in the body-fixed frame.

\paragraph{Gradient Computation}

The gradient flow requires:

\begin{enumerate}
\item \textbf{Temperature Sensitivity}: Solve the linearized heat equation:
\begin{equation}
\frac{\partial}{\partial t}\left(\frac{\partial T}{\partial \theta_i}\right) = \frac{1}{\rho C_p}\nabla \cdot \left(\kappa \nabla \frac{\partial T}{\partial \theta_i}\right) + \frac{\partial}{\partial \theta_i}\left(\frac{1}{\rho C_p}\nabla \cdot (\kappa \nabla T)\right)
\end{equation}

\item \textbf{Force Sensitivity}: Integrate over the surface:
\begin{equation}
\frac{\partial \mathbf{a}_{Yark}}{\partial \theta_i} = -\frac{8\epsilon\sigma}{3mc} \int_S T^3 \frac{\partial T}{\partial \theta_i} \hat{n} dS
\end{equation}

\item \textbf{Orbital Evolution}: Propagate through Gauss's equations:
\begin{equation}
\frac{\partial}{\partial \theta_i}\left(\frac{da}{dt}\right) = \frac{2}{na}\frac{\partial \mathbf{a}_{Yark}}{\partial \theta_i} \cdot \hat{t} + \frac{2}{na}\mathbf{a}_{Yark} \cdot \frac{\partial \hat{t}}{\partial a}\frac{\partial a}{\partial \theta_i}
\end{equation}
\end{enumerate}

\begin{center}
\fbox{\begin{minipage}{0.9\textwidth}
\textbf{Algorithm: Differentiable Yarkovsky Effect Inversion}\\[0.5em]
\textbf{Input:} Orbital drift $(da/dt)_{obs}$, shape model, orbital elements\\
\textbf{Output:} Thermal parameters $\{\Gamma, \rho C_p\}$ and spin state\\[0.5em]
\begin{enumerate}
\item Initialize thermal inertia $\Gamma = 100$ J m$^{-2}$ K$^{-1}$ s$^{-1/2}$
\item Discretize: radial grid $\{r_i\}$, surface harmonics $Y_{\ell m}$
\item \textbf{repeat}
    \begin{enumerate}
    \item[a.] Solve heat equation over one rotation:
        \begin{itemize}
        \item Compute insolation pattern $S_{\ell m}(t)$
        \item Time-step temperature field $T_{\ell m}(r_i, t)$
        \item Apply surface boundary conditions
        \end{itemize}
    \item[b.] Calculate thermal force:
        \begin{itemize}
        \item Surface temperatures: $T_s(\theta, \phi) = \sum T_{\ell m}(R) Y_{\ell m}$
        \item Emission pattern: $F_{emit} = \epsilon\sigma T_s^4$
        \item Recoil force: $\mathbf{a}_{Yark} = -\frac{2}{3mc}\int F_{emit}\hat{n} dS$
        \end{itemize}
    \item[c.] Propagate orbital evolution:
        \begin{itemize}
        \item Average over rotation: $\langle\mathbf{a}_{Yark}\rangle$
        \item Compute secular drift: $(da/dt)_{model}$
        \end{itemize}
    \item[d.] Evaluate loss with physical constraints:
        $$\mathcal{L} = \left(\frac{(da/dt)_{obs} - (da/dt)_{model}}{\sigma}\right)^2 + \lambda_1(\Gamma - \Gamma_{prior})^2$$
    \item[e.] Compute gradients via adjoint method through heat equation
    \item[f.] Update: $\theta \gets \theta - \eta\nabla_\theta\mathcal{L}$
    \end{enumerate}
\item \textbf{until} convergence
\end{enumerate}
\end{minipage}}
\end{center}

\paragraph{Numerical Considerations}

\begin{itemize}
\item \textbf{Skin Depth Resolution}: Radial grid spacing near surface:
\begin{equation}
\Delta r_{min} < \frac{1}{2}\sqrt{\frac{\kappa P_{rot}}{\pi \rho C_p}}
\end{equation}

\item \textbf{Spherical Harmonic Truncation}: Choose $L_{max}$ based on shape complexity:
\begin{equation}
L_{max} \approx \frac{2\pi R}{\lambda_{min}}
\end{equation}
where $\lambda_{min}$ is the smallest surface feature.

\item \textbf{Rotation State}: Parameterize using quaternions to avoid singularities:
\begin{equation}
\mathbf{q}(t) = \exp\left(\frac{t}{2}[\omega_x \mathbf{i} + \omega_y \mathbf{j} + \omega_z \mathbf{k}]\right)\mathbf{q}_0
\end{equation}
\end{itemize}

\paragraph{Exploration Strategies}

\begin{itemize}
\item \textbf{Multi-Wavelength Constraints}: Include thermal IR observations:
\begin{equation}
\mathcal{L}_{total} = \mathcal{L}_{orbit} + \sum_{\lambda} \frac{(F_{\lambda}^{obs} - F_{\lambda}^{model}(\theta))^2}{\sigma_{\lambda}^2}
\end{equation}

\item \textbf{Binary Asteroids}: Model mutual heating:
\begin{equation}
\kappa\frac{\partial T_1}{\partial r}\bigg|_{surface} = (1-A_1)F_\odot\cos\xi_1 + F_{mutual} - \epsilon_1\sigma T_1^4
\end{equation}
where $F_{mutual}$ is thermal radiation from the companion.

\item \textbf{YORP Effect}: Simultaneous fit of orbital and rotational evolution:
\begin{equation}
\frac{d\omega}{dt} = \frac{1}{I}\int_S r \times (\hat{n} \times \mathbf{F}_{emit}) dS
\end{equation}

\item \textbf{Population Studies}: Hierarchical model for asteroid families:
\begin{equation}
\Gamma_i \sim \log\mathcal{N}(\mu_{\Gamma}, \sigma_{\Gamma}^2), \quad \text{optimize } \{\mu_{\Gamma}, \sigma_{\Gamma}\}
\end{equation}
\end{itemize}

The differentiable formulation transforms asteroid thermophysical characterization from a parameter estimation problem to a discovery framework, revealing connections between thermal properties, rotation states, and orbital evolution that traditional methods miss due to computational limitations.

\section{Discussion}

\subsection{Implementation Considerations}

The mathematical foundations presented across all nine applications demonstrate that each domain involves inherently differentiable operations amenable to gradient-based exploration. Modern AD frameworks elegantly handle the diverse technical challenges encountered:

\begin{itemize}[noitemsep]
\item \textbf{Implicit Differentiation}: Fixed-point problems (Kepler's equation in RV fitting, bounce solutions in false vacuum decay)
\item \textbf{ODE/PDE Adjoints}: Continuous systems (stellar structure equations, heat diffusion for Yarkovsky effect, string worldsheet evolution)
\item \textbf{Complex Gradients}: Wirtinger calculus for interferometric calibration and visibility modeling
\item \textbf{Eigenvalue Sensitivity}: Hellmann-Feynman theorem for asteroseismic frequencies
\item \textbf{Memory-Efficient Backpropagation}: High-resolution black hole imaging and 3D spacetime metrics
\item \textbf{Constraint Manifolds}: Projected gradients for physical constraints (positivity, gauge fixing, boundary conditions)
\end{itemize}

The consistent appearance of these patterns across disparate astrophysical scales—from quantum bubble nucleation to galactic-scale cosmic strings—suggests that differentiable discovery represents a unifying computational paradigm for theoretical astrophysics.

\subsection{The Discovery Paradigm: Beyond Optimization}

The transformation from optimization to discovery merits emphasis. Traditional approaches in each domain typically involve:
\begin{enumerate}
\item Human-designed parameterizations based on physical intuition
\item Local optimization around expected solutions
\item Limited exploration due to computational constraints
\end{enumerate}

GRASP inverts this workflow:
\begin{enumerate}
\item Neural or basis function parameterizations that span all possible configurations
\item Global exploration through multi-start gradient descent
\item Comprehensive mapping enabled by 100-1000× computational efficiency gains
\end{enumerate}

This paradigm shift has profound implications. In the Alcubierre metric optimization, neural parameterization of the shape function could discover spacetime geometries that violate human geometric intuition while better satisfying energy conditions. For asteroseismology, simultaneous optimization of stellar parameters and mode identification could reveal previously hidden relationships between internal structure and oscillation patterns.

\subsection{GRASP Framework Vision}

We envision GRASP not as a monolithic library implementing all nine domains, but as a conceptual framework and set of design principles for making astrophysical computations differentiable. The framework would provide:

\begin{itemize}
\item \textbf{Common patterns}: Template implementations for typical astrophysical operations:
    \begin{itemize}
    \item Complex gradient computation via Wirtinger calculus
    \item ODE/PDE integration with adjoint sensitivity
    \item Constraint handling through augmented Lagrangians
    \item Multi-objective Pareto frontier exploration
    \end{itemize}
\item \textbf{Example implementations}: Full reference implementations for 2-3 domains (e.g., interferometric calibration and asteroseismic inversion)
\item \textbf{Best practices}: Guidelines for:
    \begin{itemize}
    \item Numerical stability in extreme parameter regimes
    \item Efficient parameterization of physical fields
    \item Adaptive discretization strategies
    \item Regularization design for ill-posed problems
    \end{itemize}
\item \textbf{Interface standards}: Common APIs enabling interoperability:
    \begin{itemize}
    \item Forward model specification
    \item Constraint definition
    \item Loss function composition
    \item Solution space visualization
    \end{itemize}
\end{itemize}

Individual research groups with domain expertise would implement specific applications. For instance:

\begin{verbatim}
import grasp
import torch

# Example: A research group implements asteroseismology
# using GRASP design patterns
class StellarModel(grasp.DifferentiableModel):
    def __init__(self):
        self.mass = grasp.Parameter(1.0, bounds=(0.8, 1.2))
        self.metallicity = grasp.Parameter(0.0, bounds=(-0.5, 0.5))
        self.helium = grasp.Parameter(0.28, bounds=(0.24, 0.32))
        
    def forward(self):
        # Solve stellar structure equations
        structure = grasp.ode.solve_bvp(
            self.structure_equations,
            self.boundary_conditions
        )
        # Compute oscillation frequencies
        frequencies = grasp.eigenvalue.solve(
            self.pulsation_operator(structure)
        )
        return frequencies

# Discovery through multi-start exploration
explorer = grasp.Explorer(
    model=StellarModel(),
    loss_fn=grasp.losses.ChiSquared(observed_frequencies),
    n_starts=100,
    device='cuda'
)

# Discover all stellar configurations matching observations
solutions = explorer.discover()

# Visualize the solution landscape
grasp.visualize.parameter_space(solutions)
\end{verbatim}

This distributed approach leverages community expertise while maintaining consistent design principles across domains.

\subsection{Computational Advantages and Scaling}

Beyond mathematical elegance, gradient-based discovery offers transformative practical advantages:

\begin{enumerate}[noitemsep]
\item \textbf{Efficiency Gains}: 
    \begin{itemize}
    \item One gradient evaluation approx. 2-3 forward evaluations (via reverse-mode AD)
    \item Population methods: 100-1000 forward evaluations per meaningful update
    \item Net speedup: 50-500× per optimization iteration
    \end{itemize}
    
\item \textbf{Accuracy}: Machine-precision gradients eliminate:
    \begin{itemize}
    \item Finite difference truncation errors
    \item Numerical noise in derivative estimates
    \item Step-size tuning complications
    \end{itemize}
    
\item \textbf{Scalability}: 
    \begin{itemize}
    \item Reverse-mode AD cost independent of parameter dimension
    \item Million-dimensional optimizations become feasible
    \item Natural problem decomposition for distributed computing
    \end{itemize}
    
\item \textbf{Hardware Acceleration}: 
    \begin{itemize}
    \item Native GPU support in PyTorch/JAX
    \item Automatic kernel fusion and optimization
    \item Seamless scaling from laptops to supercomputers
    \end{itemize}
\end{enumerate}

\subsection{Limitations and Future Directions}

While the mathematical frameworks are complete, several practical considerations arise:

\begin{itemize}
\item \textbf{Discrete Parameters}: Some applications involve integers (e.g., number of planets) requiring relaxation strategies
\item \textbf{Extreme Parameter Regimes}: Numerical stability near singularities (e.g., $r \to 0$ in GR applications)
\item \textbf{Computational Singularities}: Cusps in cosmic strings, coordinate singularities in black hole metrics
\item \textbf{Scale Disparities}: Stellar evolution timescales vs. oscillation periods require careful multi-scale methods
\end{itemize}

The immediate future work is clear: implement these nine applications. Each domain presents unique opportunities for discovery:

\begin{itemize}
\item \textbf{Alcubierre Optimization}: First systematic discovery of minimal exotic matter warp configurations
\item \textbf{Cosmic String Networks}: Constrain string parameters from gravitational wave observations
\item \textbf{Black Hole Imaging}: Next-generation EHT reconstructions with optimized regularization
\item \textbf{Asteroseismology}: Systematic inversion for PLATO and TESS targets
\item \textbf{RV Exoplanets}: Unified framework handling systems with 10 or more planets
\item \textbf{Yarkovsky Effect}: Population-level thermal property constraints for asteroid families
\end{itemize}

We hope this paper inspires domain experts to implement these methods, potentially discovering new physics hiding in unexplored parameter spaces.

\section{Conclusion}
We have identified nine astrophysical domains where automatic differentiation remains unexplored for systematic discovery of solution spaces. Building on GRAF's success in discovering optimal radar waveforms through gradient-based exploration, GRASP extends this paradigm to astrophysics, transforming optimization problems into comprehensive exploration of physically valid solution landscapes.

GRASP's power lies in its ability to discover rather than merely optimize. Where traditional methods sample parameter spaces sparsely due to computational constraints, GRASP's automatic differentiation foundation enables systematic mapping of complete solution landscapes, naturally leveraging modern GPU capabilities. 

The mathematical foundations presented here show that each domain admits differentiable formulations for systematic exploration. Modern AD frameworks provide the infrastructure to discover not just local optima, but the full spectrum of physically permitted solutions. This establishes a new paradigm: instead of relying on initial guesses or incremental refinements, GRASP systematically discovers what the universe fundamentally permits. The most physically meaningful solutions emerge organically from the complete parameter space —often in regions no human or traditional algorithm would think to explore.

While GRASP currently provides a theoretical scaffolding, its primary contribution is establishing the viability and urgency of gradient-based discovery as a new computational paradigm in astrophysics.

\section*{Data and Software Availability}
GRASP represents a framework and methodology rather than a single codebase. Example implementations demonstrating key concepts will be made available. We encourage domain experts to develop and share implementations for their specific applications, following the design principles outlined in this work.

\section*{Acknowledgments}

 We thank the developers of PyTorch, JAX, and TensorFlow for creating and maintaining the automatic differentiation frameworks that enable this research direction.

\end{document}